\documentclass[aps,prb,amsfonts,amsmath,amssymb,twocolumn,superscriptaddress,floatfix]{revtex4-2}
\usepackage{bm}
\pdfoutput=1
\usepackage[T1]{fontenc}
\usepackage[latin1]{inputenc}
\usepackage{dcolumn,graphicx,color,booktabs,microtype,afterpage} 
\usepackage[charter,greekuppercase=italicized]{mathdesign}
\usepackage{sidecap}
\usepackage{amsmath}
\usepackage[colorlinks,plainpages=false,linkcolor=blue,urlcolor=blue,citecolor=blue,pdfpagemode=UseNone,pdfstartview=FitBH]{hyperref}
\graphicspath{{Figures/}}
\newcommand{\cusnoh}{CuSn(OH)$_6$}
\newcommand{\cusnod}{CuSn(OD)$_6$}
\newcommand{\casnoh}{CaSn(OH)$_6$}
\newcommand{\mgsnoh}{MgSn(OH)$_6$}

\begin{document}
%
%

\title{Correlated proton disorder in the crystal structure of the double hydroxide  perovskite \cusnoh}

\author{Anton\,A.\,Kulbakov}
\affiliation{Institut\,f{\"u}r\,Festk{\"o}rper-\,und\,Materialphysik,\,Technische\,Universit{\"a}t\,Dresden,\,01069\,Dresden,\,Germany}

\author{Ellen\,H\"au\ss ler}
\affiliation{Fakult{\"a}t\,f{\"u}r\,Chemie\,und\,Lebensmittelchemie,\,Technische\,Universit{\"a}t\,Dresden,\,01062\,Dresden,\,Germany}

\author{Kaushick\,K.\,Parui}
\affiliation{Institut\,f{\"u}r\,Festk{\"o}rper-\,und\,Materialphysik,\,Technische\,Universit{\"a}t\,Dresden,\,01069\,Dresden,\,Germany}

\author{ Aswathi\,Mannathanath\,Chakkingal  }
\affiliation{Institut\,f{\"u}r\,Festk{\"o}rper-\,und\,Materialphysik,\,Technische\,Universit{\"a}t\,Dresden,\,01069\,Dresden,\,Germany}

\author{ Nikolai\,S.\,Pavlovskii }
\affiliation{Institut\,f{\"u}r\,Festk{\"o}rper-\,und\,Materialphysik,\,Technische\,Universit{\"a}t\,Dresden,\,01069\,Dresden,\,Germany}
	
\author{Vladimir\,Yu.\,Pomjakushin }
\affiliation{Laboratory\,for\,Neutron\,Scattering\,and\,Imaging\,(LNS),\,Center\,for\,Neutron\,and\,Muon\,Sciences\,(CNM),\,PSI,\,CH-5232\,Villigen,\,Switzerland}	

\author{ Laura\,Ca\~{n}adillas-Delgado }
\affiliation{Institut\,Laue-Langevin,\,BP\,156,\,38042\,Grenoble\,Cedex\,9,\,France}

\author{ Thomas\,Hansen }
\affiliation{Institut\,Laue-Langevin,\,BP\,156,\,38042\,Grenoble\,Cedex\,9,\,France}

\author{Darren\,C.\,Peets }
\affiliation{Institut\,f{\"u}r\,Festk{\"o}rper-\,und\,Materialphysik,\,Technische\,Universit{\"a}t\,Dresden,\,01069\,Dresden,\,Germany}

\author{Thomas\,Doert}\email[Corresponding author: ]{thomas.doert@tu-dresden.de}
\affiliation{Fakult{\"a}t\,f{\"u}r\,Chemie\,und\,Lebensmittelchemie,\,Technische\,Universit{\"a}t\,Dresden,\,01062\,Dresden,\,Germany}
	
\author{Dmytro\,S.\,Inosov }\email[Corresponding author: ]{dmytro.inosov@tu-dresden.de}
\affiliation{Institut\,f{\"u}r\,Festk{\"o}rper-\,und\,Materialphysik,\,Technische\,Universit{\"a}t\,Dresden,\,01069\,Dresden,\,Germany}
\affiliation{W\"urzburg-Dresden\,Cluster\,of\,Excellence\,on\,Complexity\,and\,Topology\,in\,Quantum\,Matter--ct.qmat,\,Technische\,Universit{\"a}t\,Dresden,\,01069\,Dresden,\,Germany}

\begin{abstract}
\cusnoh~is a quantum spin system from the family of magnetic double perovskite hydroxides, having a frustrated magnetic sublattice. It is also known as the natural mineral mushistonite, whose crystal structure has remained elusive for decades. Here we employ x-ray and neutron powder diffraction to solve the crystal structure of \cusnoh~and propose a structure model in the orthorhombic space group $Pnnn$ with correlated proton disorder. The occupation of the hydrogen sites in the structure is constrained by ``ice rules'' similar to those known for water ice, albeit with only local correlations. The resulting frustration of the hydrogen bonding network is likely to have a complex and interesting interplay with the strong magnetic frustration expected in the face-centred magnetic sublattice. Structural distortions, which are quite pronounced in Cu$^{2+}$ compounds due to the Jahn-Teller effect, partially alleviate both types of frustration. We also show that hydrostatic pressure tends to suppress proton disorder through a sequence of proton-ordering transitions, as some of the split hydrogen sites merge already at 1.75\,GPa while others show a tendency toward possible merging at higher pressures.
\end{abstract}

\maketitle

\section{Introduction}

Complex perovskite oxides with transition-metal cations have been a major focus of research in condensed matter physics for decades in fields as diverse as high-$T_\text{c}$ superconductivity, spintronics, and energy conversion and catalysis, to name just a few. In particular, ordered double perovskites with two different metal cations at the $B$ site have recently attracted much attention as a promising route to develop materials with desired magnetic functionalities for advanced technological applications\,\cite{Vasala2015,Basavarajappa2022}. Double perovskites have also attracted considerable attention from the point of view of fundamental research in magnetism, mainly because strongly correlated 3$d$ ions with almost completely quenched orbital moments can coexist with strongly spin-orbit-coupled 5$d$ ions in a $B$-site-ordered double-perovskite structure, resulting in strong exchange anisotropy and versatile magnetic behavior\,\cite{Morrow2015, Gangopadhyay2016, Morrow2016,Morrow2013,Manna2016, Terzic2017, Bhowal2018, Morrow2018, Aczel2019,Revelli2019,Jin2022}. Furthermore, a rock-salt-like arrangement of magnetic and non-magnetic transition-metal ions on the $B$ site results in a face-centered cubic (fcc) magnetic sublattice\,---\,a realization of one of the simplest geometrically frustrated 3D magnetic lattices consisting of regular edge-sharing tetrahedra\,\cite{Cook2015}.

In the face of such attention to perovskite oxides, it is striking that a whole class of closely related hydroxides\,\cite{NielsonKurzman11, MitchellWelch17} remain very little studied, both in terms of their crystal chemistry and their low-temperature physical properties. Hydroxide perovskites (hydroxyperovskites) are often discussed as $A$-site-vacant perovskites with the general composition {\large$\square$}\,$B^{3+}$(OH)$_3$ (single perovskites) or {\large$\square$}$_2$($BB^{\prime\,}$)(OH)$_6$ (double perovskites), where $B$ and $B^\prime$ are transition metals and {\large$\square$} denotes the vacant $A$ site. An alternative classification has been proposed by Bock and M\"uller\,\cite{BockMueller02}, who instead assign such hydroxides to the ReO$_3$ aristotype structure. The ReO$_3$ structure type with space group $Pm\overline{3}m$ has an undistorted three-dimensional network of corner-sharing octahedra with Re--O--Re bond angles of 180\textdegree. As for perovskites, distorted variants and ordered substitution variants of the ReO$_3$-type structure are numerous\,\cite{EvansWu20}. 

All known hydroxyperovskites crystallize with lower symmetry due to tilts of the (OH)$_6$  octahedra, which fall in the very limited range of 14\textdegree--17\textdegree\, tilt angles \cite{Najorka2019}. The highest symmetry, namely space group $Im\overline{3}$ for a $B$(OH)$_3$ compunds is found for the Sc(OH)$_3$ structure type, which is also adopted by the high-temperature modification of Ga(OH)$_3$, for In(OH)$_3$ and Lu(OH)$_3$ \cite{Schubert1948, Mullica1979, Mullica1980, WelchKleppe16}. Note that the $I$-centered unit cell of the structure already includes a doubling of the cubic lattice parameter with restpect to ReO$_3$. 
The tilt systems and space groups of hydroxyperovskites can be assorted to the generally accepted classification for perovskites\,\cite{Glazer1972, Howard2003, Howard2004, MitchellWelch17}. Compounds of the Sc(OH)$_3$ type belong to the $a^+a^+a^+$ tilt system. 
Ordered double hydroxide perovskite of the composition ($BB^{\prime\,}$)(OH)$_6$ and local site or electronic constraints, e.g. for Jahn-Teller active cations, require additional symmetry reductions.


Hydroxide perovskites are characterized by the presence of correlated proton disorder in their lattice, which has only recently been demonstrated in several members of this family\,\cite{LafuenteYang15, WelchKleppe16,Kampf2024,Basciano98}. The phenomenon of correlated proton disorder is best known in conventional water ice (hexagonal\,ice,\,I$_\text{h}$)\,\cite{BernalFowler33, Pauling35, Malenkov09}. In the absence of long-range order, the proton network in H$_2$O ice has an exponentially large number of ground-state configurations that obey so-called ``ice rules''\,\cite{BernalFowler33, Pauling35}. Each oxygen atom forms two short covalent bonds and two longer hydrogen bonds with adjacent protons. This ``two-in, two-out'' configuration results in a highly correlated proton network with residual entropy and complex non-Arrhenius proton dynamics, requiring collective quantum tunneling between different proton configurations on hexagonal plaquettes\,\cite{CastroNeto06, BentonSikora16}, giving rise to a telltale peak in quasielastic neutron scattering\,\cite{BoveKlotz09, Komatsu22}.

Of course, water ice is not the only compound to exhibit correlated proton disorder. It is a rather general phenomenon for H$^+$ ions involved in hydrogen bonding, because in the O--H$\,\cdots$O (or similarly N--H$\,\cdots$O) configuration the proton sits in a double-well potential with the possibility of tunneling between the equivalent O--H$\,\cdots$O and O$\,\cdots$H--O configurations. As the actual hydrogen position cannot be determined by x-ray with the required accuracy, the respective crystal structures are often reported either with split atom positions with partial occupation \,\cite{Viswanathan18} or even completely without proton positions, especially in older works \,\cite{Schubert1948}. However, the number of \textit{magnetic} materials which exhibit correlated proton disorder, e.g. in metal-organic frameworks, remains rather limited, and the interplay of such disorder with magnetism is a topic of current interest in the context of molecular multiferroics\,\cite{JainRamachandran09, HuKurmoo09, CanadillasDelgado12, MaczkaGagor14, ViswanathanBhat19, Viswanathan19}. It would certainly be interesting to study correlated proton disorder in the context of frustrated magnetism, but we are not aware of any such studies.

Ga(OH)$_3$ (s\"ohngeite)\,\cite{WelchKleppe16} and MnSn(OH)$_6$ (tetrawickmanite)\,\cite{LafuenteYang15} are two examples for which split-atom models with pairs of half-occupied hydrogen positions have been reported based on combined x-ray and neutron diffraction with Raman spectroscopy. A common feature of these structure models are 4-membered oxygen rings with partially occupied hydrogen positions located between the O atoms. The case of s\"ohngeite exemplarily highlights another problem: from combined single-crystal x-ray diffraction data and Raman spectroscopy, it was impossible to unambiguously determine the correct space group of the ambient-temperature polymorph, as the crystal structure of Ga(OH)$_3$ can equally well be described with space groups $P4_2/nmc$ (no.~137) and $P4_2/n$ (no.~86) \,\cite{WelchKleppe16}. Ga(OH)$_3$ undergoes a phase transition to a cubic structure with $Im\overline{3}$ (no.~204) symmetry on heating to 423\,K. However, Raman spectroscopy has been successfully used to distinguish between two different structural models of stottite, FeGe(OH)$_6$ \cite{Kleppe2012}. The structural transition to the cubic phase has been associated with the rearrangement of protons from the 4-membered rings mentioned above into ``crankshafts''\,\cite{WelchKleppe16}. 

A very similar hydrogen bonding network with correlated proton disorder has also been found in the double hydroxide perovskite MnSn(OH)$_6$, which is also known to crystallize in both cubic (wickmanite, $Pn\overline{3}$) and tetragonal (tetrawickmanite, $P4_2/n$ or $P4_2/nnm$) polymorphs. Four out of five hydrogen positions in the tetrawickmanite structure are statistically disordered. Due to the strong Coulomb repulsion, the two adjacent sites in the 4-membered ring cannot be occupied simultaneously, resulting in a correlation between the proton positions such that each ring can be oriented in a homodromic arrangement, either clockwise or anticlockwise. This clockwise-or-anticlockwise rule is analogous to the ice rules in water ice in the sense that it allows for macroscopic degeneracy and residual entropy, and it is natural to expect that quantum tunneling between these two homodromic proton configurations should be possible, which would lead to nontrivial proton dynamics in these compounds.

It remains an open question whether the described proton disorder is a common feature of all hydroxide perovskites or just a peculiarity of the mentioned members of this family. Most of the other published crystal structure refinements were based only on x-ray diffraction (XRD) data. Although some structure models may contain split positions for the hydrogen atoms, (correlated) proton disorder was not explicitly discussed until 2015\,\cite{LafuenteYang15}. However, there is no obvious reason to expect that the structures of Ga(OH)$_3$ and MnSn(OH)$_6$ should be somehow special in this abundant family of compounds. On the other hand, it has long been known that hydroxide perovskites are good proton conductors above room temperature\,\cite{Jena2004}, suggesting that protons in these lattices are highly mobile. Our present results show that correlated proton disorder is also inherent in the structure of CuSn(OH)$_6$. It seems plausible that this may be a general feature of many, if not all, hydroxide perovskites, which would make them an ideal platform for studying the influence of correlated disorder and proton ordering transitions on frustrated magnetism.

\section{Experimental details}

\subsection{Sample preparation} \label{SubSec:synthesis}

\begin{figure}[b]
	\includegraphics[width=1.0\linewidth]{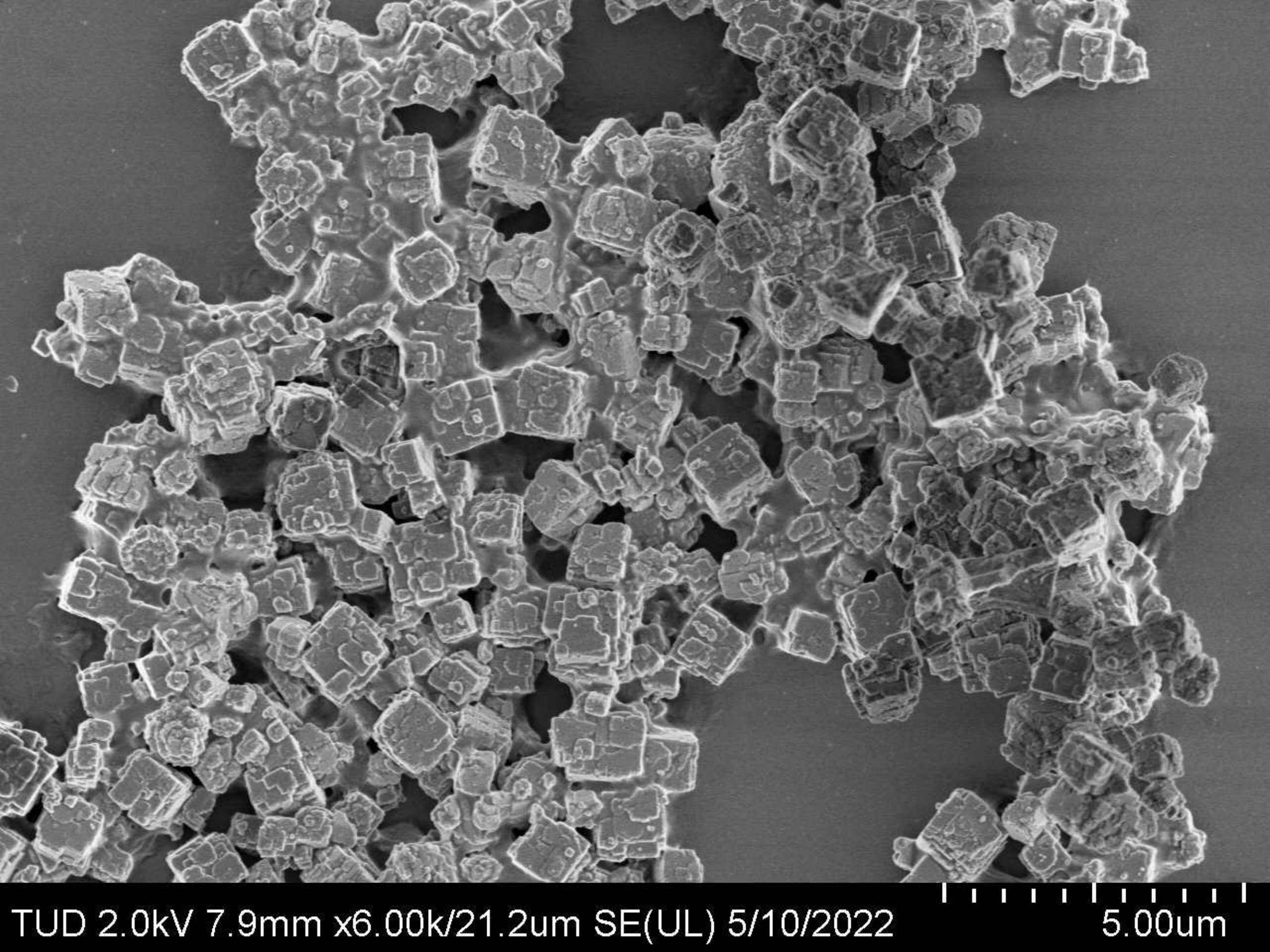}
	\caption{An SEM image of the \cusnoh~powder sample.}
	\label{fig:SEM}
\end{figure}

We have synthesized phase-pure powder samples of CuSn(OH)$_6$ by co-precipitation from aqueous solutions of CuCl$_2\cdot2$H$_2$O and Na$_2$Sn(OH)$_6$ at ambient temperature. Equimolar amounts of the starting materials were ground together prior to dissolution in water (with $c = 0.16$\,mol/L). Grinding appeared to be beneficial in achieving a phase-pure product. After several days of stirring, the precipitate was allowed to settle, filtered, washed with ethanol, and dried at ambient temperature in a vacuum drying chamber.
Depending on the batch size a slight color change (from turquoise to white-pale blue) is visible during the first days, indicating the consumption of [Cu(H$_2$O)$_6$]$^{2+}$ ions. To ensure the complete turnover of the starting materials, the reaction mixture was stirred for two additional days after the color change. Longer or more vigorous stirring resulted in visible amounts of polytetrafluoroethylene (PTFE, see below), which originate from slight abrasions of the PTFE coated magnetic stirring bar. However, as the contamination is found on top of the reaction mixture, it could be decanted before filtering and washing of the target phase.
The deuterated version of the same compound, CuSn(OD)$_6$, was synthesized by the same procedure from a solution in D$_2$O. The resulting pale blue powders contain intergrown $\mu$m-sized cubic crystallites (see Fig.\,\ref{fig:SEM}).

\subsection{Scanning electron microscopy (SEM)}
\label{SubSec:SEM}
SEM images were taken on a Hitachi SU8020 microscope equipped with a field emission cathode and a triple detector system for secondary and low-energy backscattered electrons ($U_\text{a}=2$\,kV, $I=5$\,$\mu$A). The samples were placed on carbon pads and sputtered with a thin gold layer before the measurements.

\subsection{X-ray and neutron powder diffraction}
\label{SubSec:XRD_PND}

\begin{figure}[t]
	\includegraphics[width=1.0\linewidth]{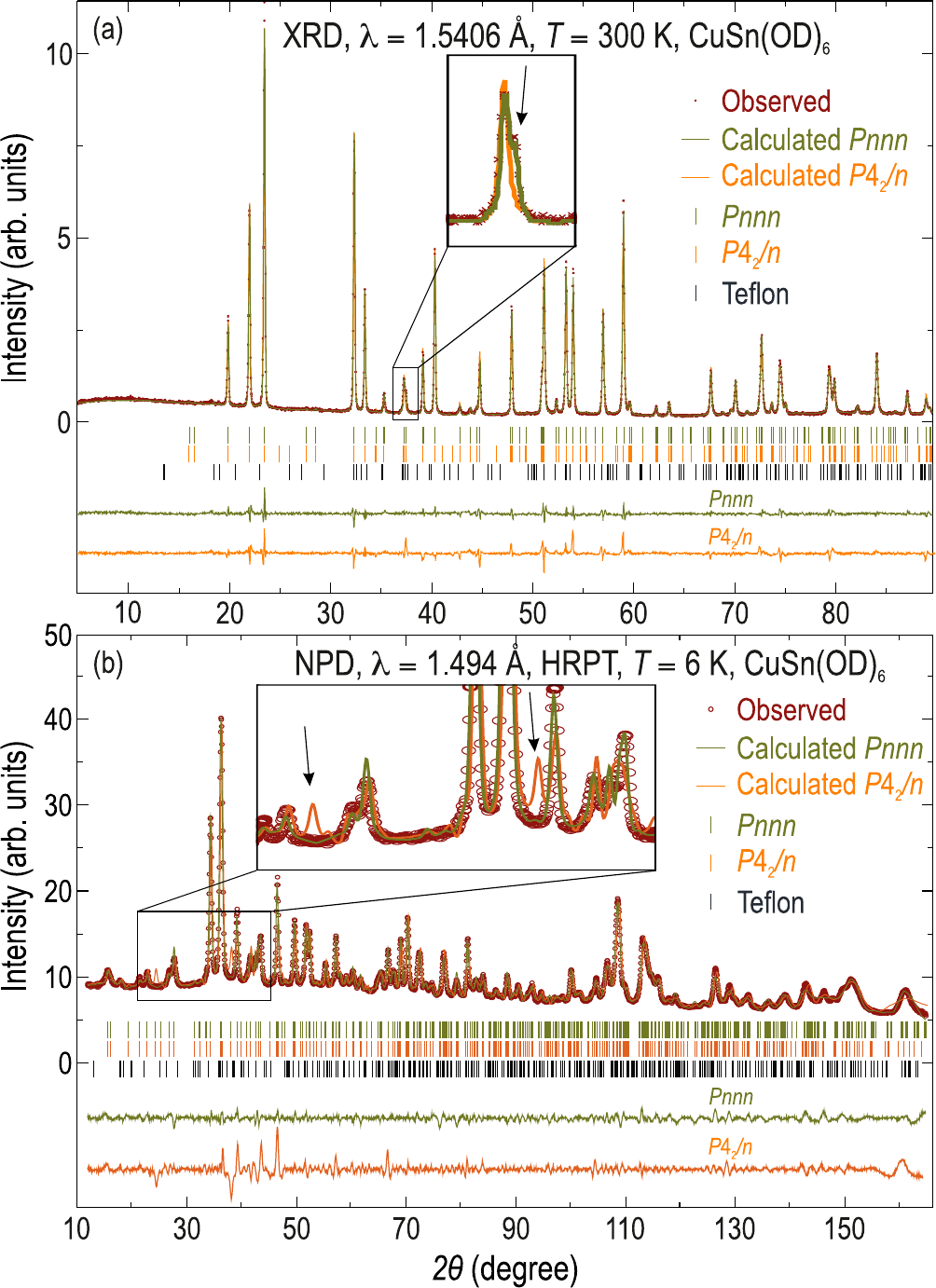}
	\caption{Scattered (a) x-ray and (b) neutron intensities at 300 and 6\,K, respectively, as a function of 2$\theta$, refined in the tetragonal $P4_{2}/n$ and orthorhombic $Pnnn$ space groups. Insets highlight regions that cannot be adequately described within $P4_{2}/n$.}
	\label{fig:fit}
\end{figure}


The quality of the deuterated \cusnod~sample was checked by x-ray diffraction measurements in Bragg-Brentano geometry at ambient conditions on an Empyrean diffractometer (Malvern Panalytical). The diffractometer is equipped with a Johannsson type Ge(111) monochromator using Cu-K${\alpha}_{1}$ radiation ($ \lambda = 1.5406$~\AA), a variable-divergence slit to keep the illuminated sample area constant, and a PIXcel3D area detector. The sample was prepared on a zero-background sample holder made from off-cut monocrystalline silicon. The powder x-ray diffractograms were compared with powder and single-crystal reference data from the PDF-, ICSD-, and in-house databases. 

 High-intensity neutron powder diffraction was measured with a sample encapsulated in a vanadium can on the high-resolution thermal-neutron powder diffractometer HRPT\,\cite{Fischer2000} at the Paul Scherrer Institute (PSI) in Villigen, Switzerland. We collected data between 2.50\textdegree\ and 164.85\textdegree\ in 0.050\textdegree\ steps, using three wavelengths: 1.494, 1.886 and 2.449\,\AA. The counting times were 6.5, 6 and 26\,h, respectively, at 6\,K. We selected the wavelength using a Ge ($HKK$) wafer-type monochromator, 28\,cm high, with variable vertical focusing and a total mosaic halfwidth~of~$15'$.
 
 In addition, to determine the pressure dependence of the hydrogen positions in the crystal structure, we performed a neutron powder diffraction experiment using a Paris-Edinburgh pressure cell\,\cite{Besson1995, Besson1995a} at D20\,\cite{Hansen2008} at the Institut Laue-Langevin (ILL), Grenoble, France. The experiment was performed at pressures from 0 to 6\,GPa using a wavelength of 1.544~\AA. The wavelength was selected using a germanium monochromator with the $(115)$ reflection. Pure Pb powder was added as an internal pressure standard.

\section{Results and Conclusions}

The first attempt to refine the crystal structure of \cusnoh~dates back to 1976\,\cite{Morgenstern1976}, but it has been emphasized by other groups \,\cite{WelchKleppe16, MitchellWelch17, Najorka2019} that the originally proposed tetragonal structure with the space group $P4_2/nnm$ (no.~134) is implausible because it involves a tilt system $a^+a^+c^0$ with a zero tilt. Therefore, the low-symmetry subgroup $P4_2/n$ (no.~84) with the tilt system $a^+a^+c^-$ was conjectured\,\cite{Najorka2019} by analogy with MnSn(OH)$_{6}$\,\cite{LafuenteYang15}. In a series of solid solutions Cu$_{1-x}$Zn$_x$Sn(OH)$_\text{6}$, a transition to the cubic ($Pn\smash{\overline{3}}$) (no.~201) structure was observed at $x\approx50$\%\,\cite{Najorka2019}. 

\begin{figure}[t]
	\includegraphics[width=0.45\textwidth]{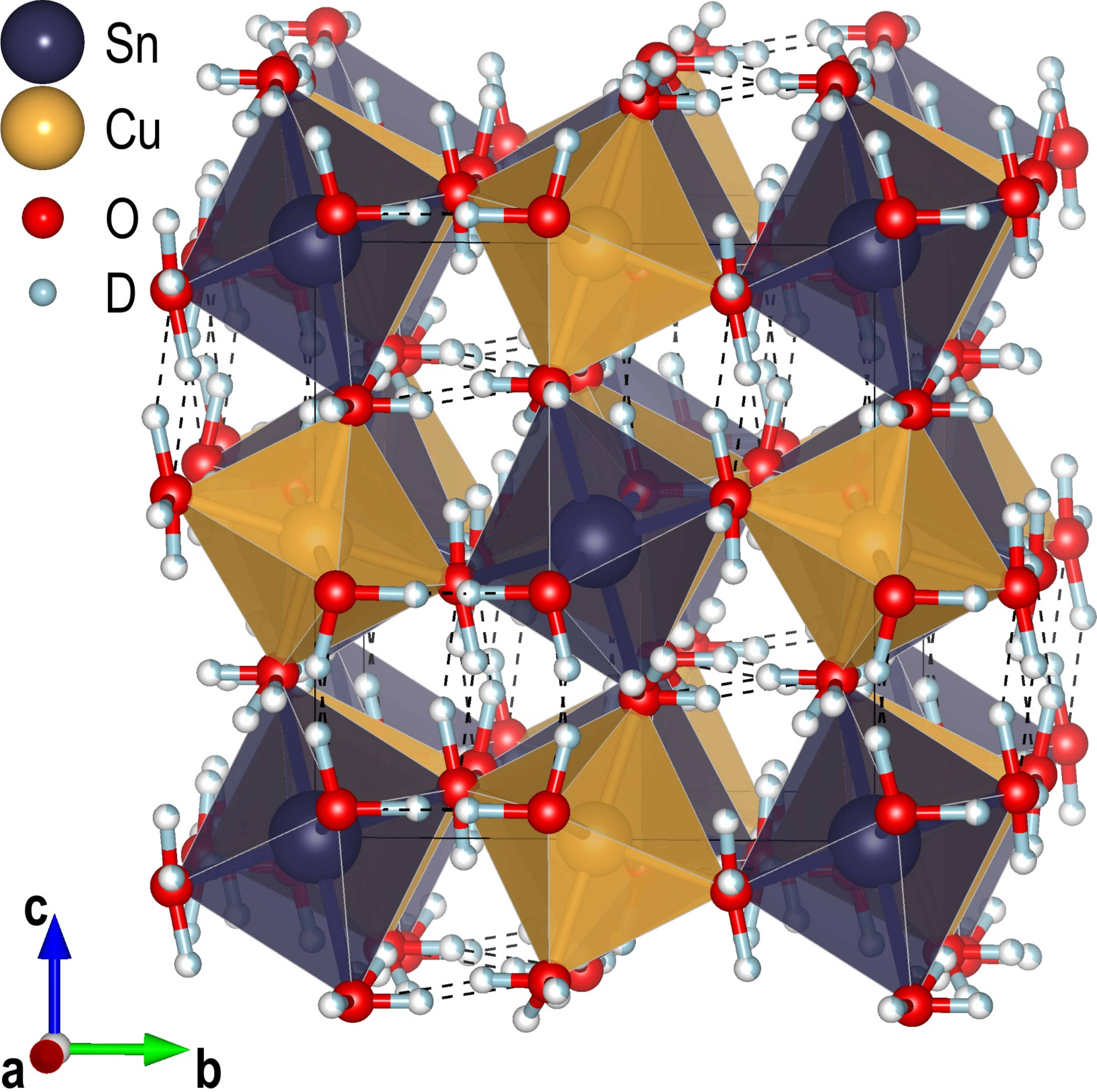}\bigskip\\
	\includegraphics[width=0.45\textwidth]{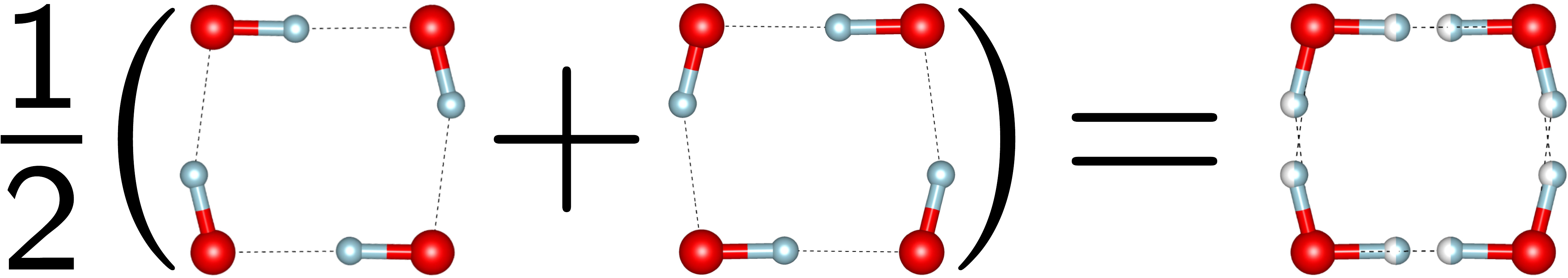}\vspace{2pt}
	\caption{\footnotesize Proposed crystal structure of \cusnod, at ambient conditions in the orthorhombic space group $Pnnn$ based on neutron diffraction measurements using split atom positions with partial occupation for hydrogen/deuterium; the total occupation of each split position is equal to 1. Accordingly, each oxygen atom is bonded to two or more partially occupied hydrogen/deuterium atoms. The hydrogen positions facing each other cannot be occupied simultaneously, which introduces correlations in the hydrogen bonding network, as schematically shown in the bottom panel. The visualisation was done in VESTA\,\cite{VESTA}.}
	\label{fig:struct}
\end{figure}

\begin{figure}
	\includegraphics[width=0.95\linewidth]{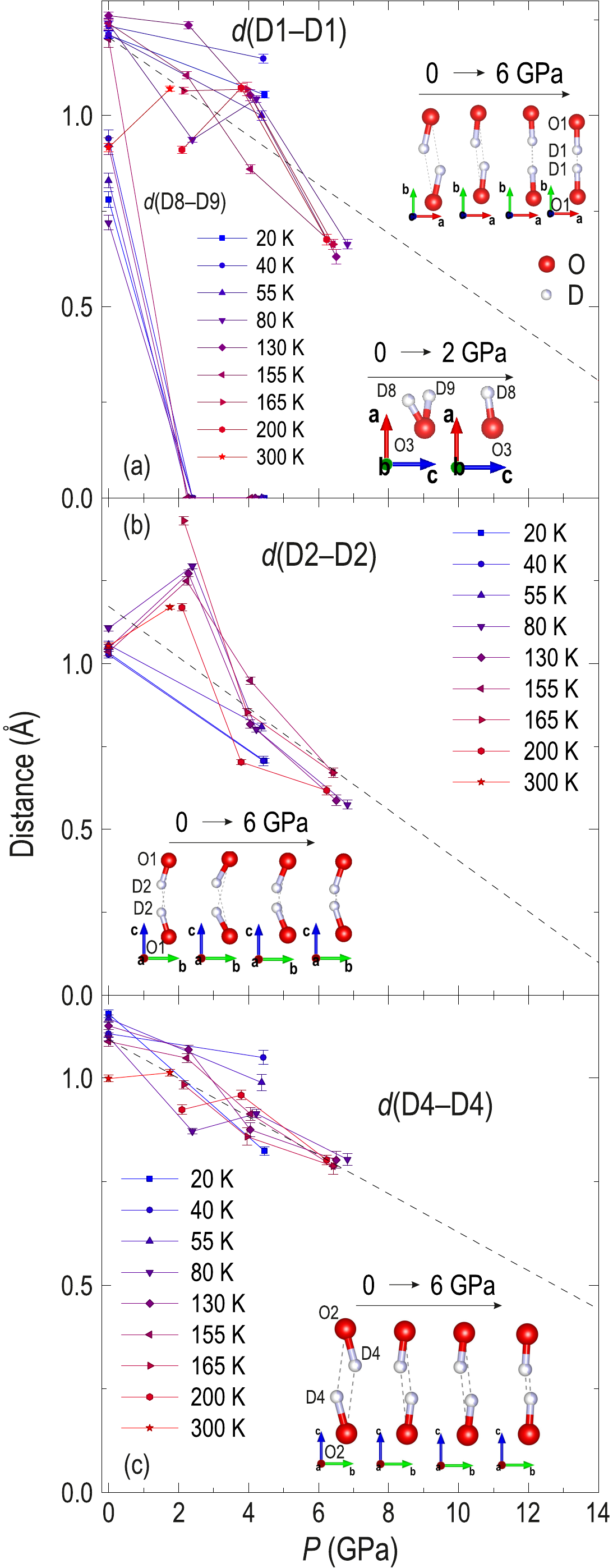}
	\caption{Pressure dependence of the distances between selected split H/D sites	in the crystal structure of deuterated	mushistonite \cusnod, obtained from the refinement of our D20 data. The insets show the evolution of the geometry of the O--D bonds.}
	\label{fig:pressure}
\end{figure}

We collected x-ray and neutron powder diffraction data on the protonated and deuterated samples, respectively, as shown in Fig.\,\ref{fig:fit} and Appendix\,\ref{appA}. Although the lattice parameters are largely consistent with tetragonal symmetry within experimental uncertainty, the room-temperature x-ray data (insensitive to hydrogen positions) and neutron data did not agree with the conjectured space groups $P4_2/nnm$\,\cite{Morgenstern1976} and $P4_2/n$\,\cite{Najorka2019}. These space group models cannot describe the peaks observed at positions $37.46^{\circ}$ for x-ray data and $24.21^{\circ}$ and $38.98^{\circ}$ for neutron data, as highlighted with arrows in the insets of Fig.\,\ref{fig:fit}.

Instead, we were able to refine a structural model in the orthorhombic space group $Pnnn$ (no.~48) with lattice parameters $a=7.5764(1)$~\AA, $b=7.5740(1)$~\AA, and $c=8.0691(1)$~\AA, as shown in Fig.\,\ref{fig:struct}. Note that the splitting between the lattice parameters $a$ and $b$ is small; the symmetry lowering might primarily result from the hydrogen bonding network. The $Pnnn$ model for \cusnoh\ implies a tilt system $a^+a^+a^+$, which can be rationalized by symmetry considerations.

\begin{figure}[t]
	\includegraphics[width=1.0\linewidth]{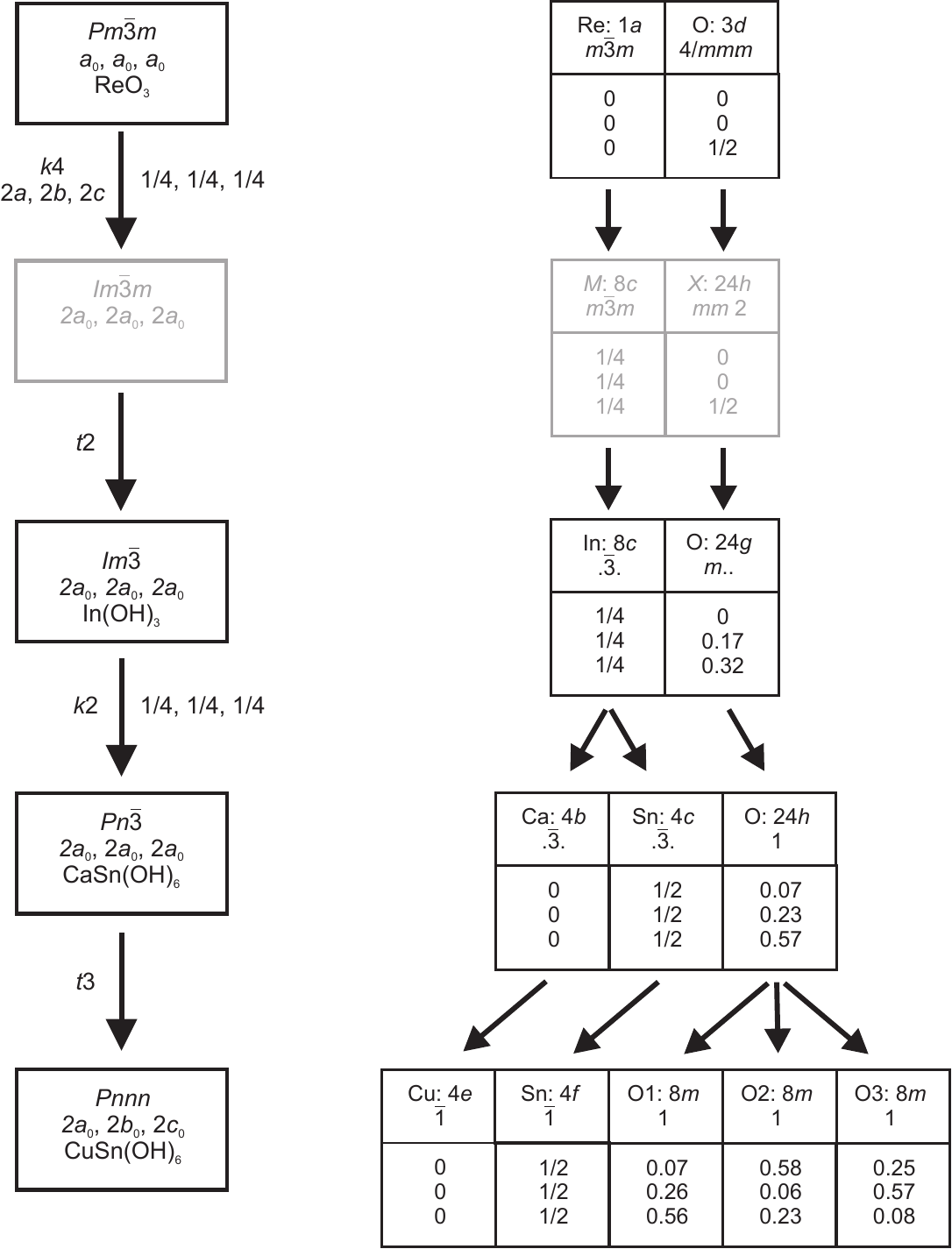}
	\caption{B\"arnighausen tree relating the crystal structure of \cusnoh\ to that of ReO$_3$; only metal and oxygen/anion positions are listed. The grey space group lacks an example, see text for details.}
	\label{fig:Baernighausen}
\end{figure}

$Pnnn$ is a maximal subgroup of \textit{Pn}$\overline{3}$, the space group of schoenfliesite, \mgsnoh~\cite{Basciano98} and burtite, \casnoh~\cite{Cohen68, Basciano98}. As these compounds share the same composition and Mg$^{2+}$ and Cu$^{2+}$ have almost identical ionic radii (0.86 vs. 0.87\,\AA, both for octahedral coordination \cite{Shannon76}), a close structural relationship between the compounds is not surprising. However, Cu$^{2+}$ exhibits a considerable Jahn-Teller distortion, which necessitates an additional degree of freedom, i.e. an additional symmetry reduction, to avoid $\overline{3}$ or $3$ site symmetry for the metal atoms. Without an enlargement of the unit cell, this additional symmetry reduction must be of the \textit{translationengleiche} type, and \textit{Pn}$\overline{3}$, $Pnnn$ is the only maximal subgroup of \textit{Pn}$\overline{3}$ to achieve this.
Indeed, it is possible to trace the space groups and atomic positions (of metal and oxygen atoms) of \cusnoh\ back to those of the ReO$_3$ aristotype using a B\"arnighausen diagram\,\cite{Baernighausen80, BockMueller02} via schoenfliesite/burtite and In(OH)$_3$, space group \textit{Im}$\overline{3}$ \cite{MitchellWelch17, Mullica1979}, with one intermediate step in space group \textit{Im}$\overline{3}{m}$, as shown in Fig.\,\ref{fig:Baernighausen}. The absence of an $MX_3$ example in space group \textit{Im}$\overline{3}{m}$ (shown in grey in the figure) is understandable, as there is no gain in structural flexibility compared to the aristotype. Notably, all distorted structures of this B\"arnighausen tree\,---\,those with lower symmetry relative to the aristotype\,---\,belong to the same $a^+a^+a^+$ tilt system.

The structure of \cusnod\ consists of alternating [Cu$^{2+}$(OD)$_{6}$] and [Sn$^{4+}$(OD)$_{6}$] octahedra connected in an angular cross-linking pattern. The [Cu$^{2+}$(OD)$_{6}$] octahedra are located at Wyckoff position $4e$, while the [Sn$^{4+}$(OD)$_{6}$] octahedra are located at Wyckoff position $4f$. The distances between Cu and O atoms are 1.943(4) (2$\times$), 2.053(4) (2$\times$) and 2.295(3)\,\AA\ (2$\times$), while those between Sn and O atoms are 2.075(3) (2$\times$), 2.079(4) (2$\times$) and 2.101(4)\,\AA\ (2$\times$). The distortion of the [Cu$^{2+}$(OD)$_{6}$] octahedra in this compound is significant, which can be attributed to the Jahn-Teller effect mentioned above. The structure contains three nonequivalent O atoms, all deuterated as OD groups and located in $8m$~Wyckoff positions, forming 4-membered rings, as shown in Fig.\,\ref{fig:struct} in the bottom panel. 

\begin{table}[h!]
	\centering
	\caption{Temperatures, unit cell parameters and $R$ factors for our refinement of x-ray and neutron diffraction data at $T=300$ and 6\,K, respectively, where $\lambda_{1}=1.494$, $\lambda_{2}=1.886$~and $\lambda_{3}=2.449$\,\AA.\smallskip}
	\label{table:Rs}
	\begin{tabular}{  l l c c c} 
   \toprule\toprule
		Parameters & X-ray&     \multicolumn{3}{c}{Neutrons}  \\
	\midrule
		Wavelength & Cu-K${\alpha}_{1}$&     $\lambda_{1}$  & $\lambda_{2}$ &     $\lambda_{3}$    \\
		\midrule
		Temperature & $T=300$~K&\multicolumn{3}{c}{$T=6$~K}\\
		$a$\,(\AA) &7.5918(3)&\multicolumn{3}{c}{7.5764(1)}\\
		$b$\,(\AA)& 7.5915(3)  &\multicolumn{3}{c}{7.5740(1)}\\
		$c$\,(\AA)&8.0965(1)&\multicolumn{3}{c}{8.0691(1)}\\
		$V$\,(\AA$^{3}$) & 466.63(3)&\multicolumn{3}{c}{463.04(1)}\\
		\midrule
		R-Factor(\%) &1.42& 2.15  &  2.08  &  1.82  \\
		Rp(\%) &4.85& 9.13 & 11.4 &   9.95 \\
		Rwp(\%) &6.31& 9.71 &  12.0  &   10.1   \\
	\bottomrule\bottomrule
	\end{tabular}
\end{table}

\begin{table}[h!]
	\centering
	\caption{Refined atomic positions in synthetic \cusnod~based on our x-ray diffraction data at 300\,K and our neutron powder diffraction data at 6\,K, corresponding to the structure shown in Fig.\,\ref{fig:struct}. Wy is the Wyckoff position. Sn, Cu, O and D have been refined with isotropic thermal parameters. Final positions have been taken from neutron powder diffraction.\smallskip} 
	\label{table:atoms}
	\begin{tabular}{l@{~\,}c@{~~}r@{.}lr@{.}lr@{.}lr@{.}lr@{.}lr@{.}} 
		\toprule\toprule
	Site & Wy & \multicolumn{2}{c}{$x$} & \multicolumn{2}{c}{$y$} & \multicolumn{2}{c}{$z$} & \multicolumn{2}{c}{$U$}&\multicolumn{2}{c}{Occ.}\\ 
	\midrule
	Sn  &  \textit{4f}&  0 & 5 & 0 &0 & 0 & 5 & 0&090(6) &  1 & 00  \\
	Cu  &    \textit{4e} &0 & 5 & 0 &0&0& 0 & 0&090(6) &  1 & 00  \\
	O1  &  \textit{8m}  & 0 & 7610(6) & $-$0 & 0620(6) & 0 &  5656(6) & 0&090(6) &  1 & 00   \\
	O2  &   \textit{8m}  & 0&43051(5) & $-$0&2518(7) & 0&5789(6) & 0&090(6) &  1&00 \\
	O3  &    \textit{8m} & 0&41845(6) & 0&0746(6) & 0&7353(5) & 0&090(6) &  1&00   \\
	D1 &     \textit{8m} &0&774(1)& $-$0&2056(9) & 0&5692(7) & 0&090(6) &  0&50\\   
	D2  &    \textit{8m}  &0&751(3) &$-$0&0281(9) & 0&6926(9) & 0&090(6) &  0&50   \\
	D3 &    \textit{8m}   &0&3054(9) &$-$0&249(4)&  0&555(1) & 0&090(6) &  0&40  \\
	D4  &    \textit{8m} & 0&431(1)& $-$0&219(1) & 0&707(1) & 0&090(6) &  0&40   \\
	D5  &    \textit{8m}  &0&397(3) &$-$0&248(7)&  0&461(2) & 0&090(6) &  0&20  \\
	D6 &   \textit{8m}    &0&419(2) & 0&201(1) & 0&738(3) & 0&090(6) &  0&30  \\
	D7 &   \textit{8m}   & 0&272(2) & 0&050(1) & 0&742(6) & 0&090(6) &  0&30   \\
	D8 &    \textit{8m}  & 0&538(2) & 0&066(2) & 0&778(2) & 0&090(6) &  0&20   \\
	D9 &   \textit{8m}   & 0&512(3) & 0&156(3) & 0&696(2) & 0&090(6) &  0&20   \\
		\bottomrule\bottomrule
	\end{tabular}
\end{table}

The resulting crystal structure for \cusnod~and refinements are shown in Fig.\,\ref{fig:struct} and Table\,\ref{table:Rs}. Note, that the sample used for neutron diffraction contains a small amount of polytetrafluoroethylene (Teflon) as an impurity, probably from the magnetic stirrer bar. This has space group $P3_1$ and $a=b=5.6651$\,\AA, $c=19.5592$~\AA\ and $\gamma=120^\circ$\,cell parameters. The refined atomic positions of \cusnod\ are given in Table\,\ref{table:atoms}. The Rietveld and Le Bail refinements of the main and impurity phases, respectively, for the x-ray and neutron powder diffraction data were performed in {\sc JANA2006}~\cite{Jana2006} and {\sc FullProf}\,\cite{Rodr_guez_Carvajal_1993}. Oxygen atoms have short contacts to between two and four partially occupied deuterium neighbors which can be interpreted as covalent bonds; their total occupations adds up to two. Simultaneous occupation of any two sites bonded to the same oxygen is forbidden. The structure model in $Pnnn$ also forbids the simultaneous occupation of any pair of hydrogen/deuterium sites facing each other, so either O--H$\,\cdots$O or O$\,\cdots$H--O configuration is allowed. This means that the hydrogen bonding network must be correlated with all O-H bonds pointing in the same direction as shown schematically in the lower panel of Fig.\,\ref{fig:struct}. This behaviour is similar to that previously observed in Ga(OH)$_3$ and MnSn(OH)$_6$\,\cite{LafuenteYang15, WelchKleppe16}, suggesting some general feature for hydroxide perovskites.

Another remarkable feature of this structure is the high degree of frustration in the hydrogen bonding network. The symmetry-equivalent oxygen positions bridged by the \mbox{O--H$\,\cdots$O} bonds are separated by 2.562, 2.667 and 2.980~\AA\ along the $a$, $b$ and $c$ directions respectively. These distances are comparable to the O--O distances along the edges of Cu-centered (1.943--2.295\,\AA) or Sn-centered (2.075--2.101\,\AA) distorted octahedra. Consequently, for every oxygen atom shared by a pair of Cu- and Sn-centered octahedra, there are eight neighbouring oxygen atoms belonging to the same two octahedra, plus two more neighbouring oxygen atoms from adjacent octahedra (across the short diagonal of rhombic vacancies seen in Fig.\,\ref{fig:struct}), all with very small distance variations of $\pm$20\%. Consequently, each oxygen atom has ten potential hydrogen bonding partners along the O--H$\,\cdots$O pathway, but only one available proton. This is expected to lead to a large number of almost degenerate configurations of the hydrogen bonding network. Such degeneracy is a consequence of the large octahedral tilts, which bring the oxygens in opposite octahedra as close together as the edge length of a single octahedron, or even closer. 

It is worth noting that a further symmetry reduction to the monoclinic space group $P2/n$ would allow an ordered structure model without partially occupied H sites and rational $M$--O--H bond angles of about 109$^\circ$. However, this model would not solve the problem of multiple degenerate configurations for the hydrogen sites. As there is no experimental evidence for a monoclinic lattice distortion, this option was not considered for the final structure model.

In addition, we performed a pressure neutron powder diffraction experiment to study the behavior of hydrogen in our material. In \cusnoh\ each hydrogen atom statistically occupies one of the two or more possible equilibrium sites. In total, there are 9 positions for hydrogen/deuterium (D1, D2, D3, D4, D5, D6, D7, D8, D9), all of which are split at ambient pressure, and 3 inequivalent positions for oxygen (O1, O2, O3). Table\,\ref{table:atoms} shows the atomic positions at ambient pressure. 

\begin{figure}
	\includegraphics[width=1.0\linewidth]{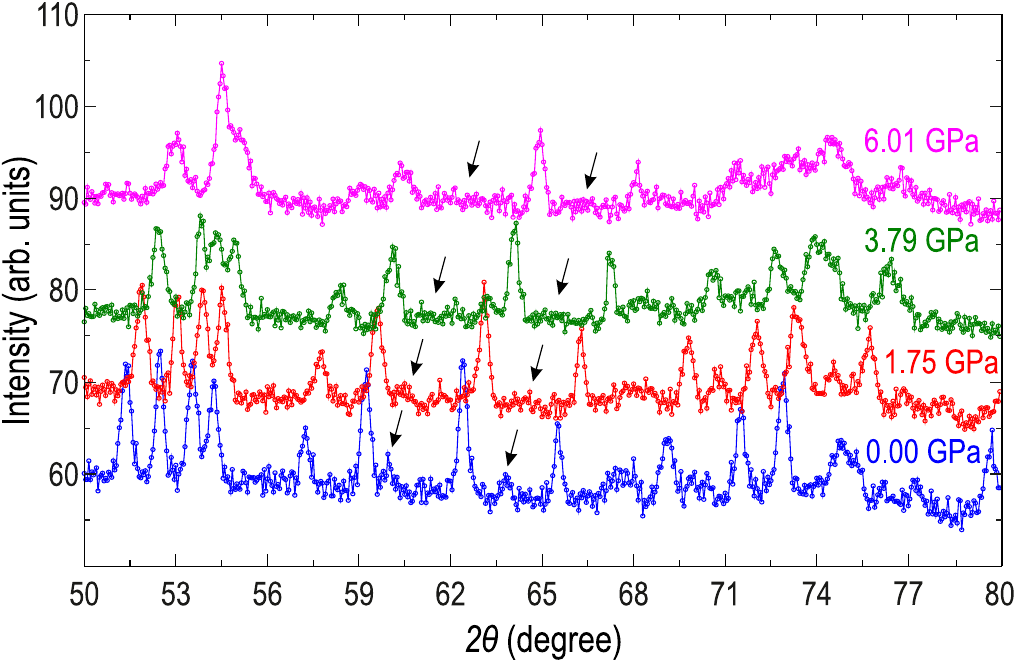}
	\caption{Scattered neutron intensities in the pressure range from 0 to 6\,GPa at 200\,K,  as a function of 2$\theta$. Black arrows indicate peaks which disappear with increasing pressure.}
	\label{fig:neutrons}
\end{figure}

Selected distances between equivalent H/D sites are shown in Fig.\,\ref{fig:pressure}. These distances were extracted from the refined crystal structures after processing the neutron data with {\sc FullProf}~\cite{Rodr_guez_Carvajal_1993}. At 1.75\,GPa, two peaks at $59.94^\circ$ and $63.80^\circ$ vanish, as shown in Fig.~\ref{fig:neutrons}.  Refinements indicate that this is caused by the merging of the D8 and D9 sites.  With three rather than four hydrogen positions bonded to O3, this should lead to a reduction in disorder in the hydrogen bonding network.  We note, however, that the merged site has the same 8m Wyckoff position as the original sites, so the symmetry of the site does not change and the space group is preserved.

The distances between the equivalent D1, D2, and D4 sites, which face each other, decrease with pressure, as shown in Fig.~\ref{fig:pressure}, although no merging occurs up to 6\,GPa. By extrapolation, we estimate that the ordering transition on these sites is likely to occur at approximately 15--20\,GPa.

We conducted measurements at several temperatures; however, as shown in Fig.\,\ref{fig:pressure}, the atomic positions and bond angles exhibit no significant temperature dependence. All extracted distances between equivalent H/D sites vary with temperature by no more than a few tenths of an {\AA}ngstr\"{o}m, which is within statistical uncertainty.

In conclusion, we present a new model for the crystal structure for \cusnod\ and its changes when hydrostatic pressure is applied. We show that the D8 and D9 sites merge into a single site at 1.75\,GPa, and other hydrogen positions show a tendency toward a possible merging around 15--20\,GPa.  The availability of a crystal structure will facilitate the determination of how hydrogen disorder affects the magnetic properties of the system. It will also allow us to determine whether this can be considered an additional factor influencing the emergence of new magnetic properties. 

\section*{Data Availability}

Samples and data are available upon reasonable request from T.\ Doert or D.\ S.\ Inosov; data underpinning this work are also available from Ref.~\onlinecite{OPARA_CuSn1}.

\begin{acknowledgments}
This project was funded by the Deutsche Forschungsgemeinschaft (DFG, German Research Foundation) through individual grants IN 209/12-1, DO 590/11-1 (Project No. 536621965), and PE~3318/2-1 (Project No.\ 452541981); through projects B03 and C03 of the Collaborative Research Center SFB~1143 (Project No.\ 247310070); and through the W\"urzburg-Dresden Cluster of Excellence on Complexity and Topology in Quantum Materials\,---\,\textit{ct.qmat} (EXC~2147, Project No.\ 390858490). The authors acknowledge the support of the Institut Laue-Langevin, Grenoble, France in providing neutron research facilities used in this work~\cite{report708}. This work is based in part on experiments performed at the Swiss spallation neutron source SINQ, Paul Scherrer Institute, Villigen, Switzerland.
\end{acknowledgments}

\appendix

\section{XRD and NPD\label{appA}}

\begin{figure}[t!]
	\includegraphics[width=1.0\linewidth]{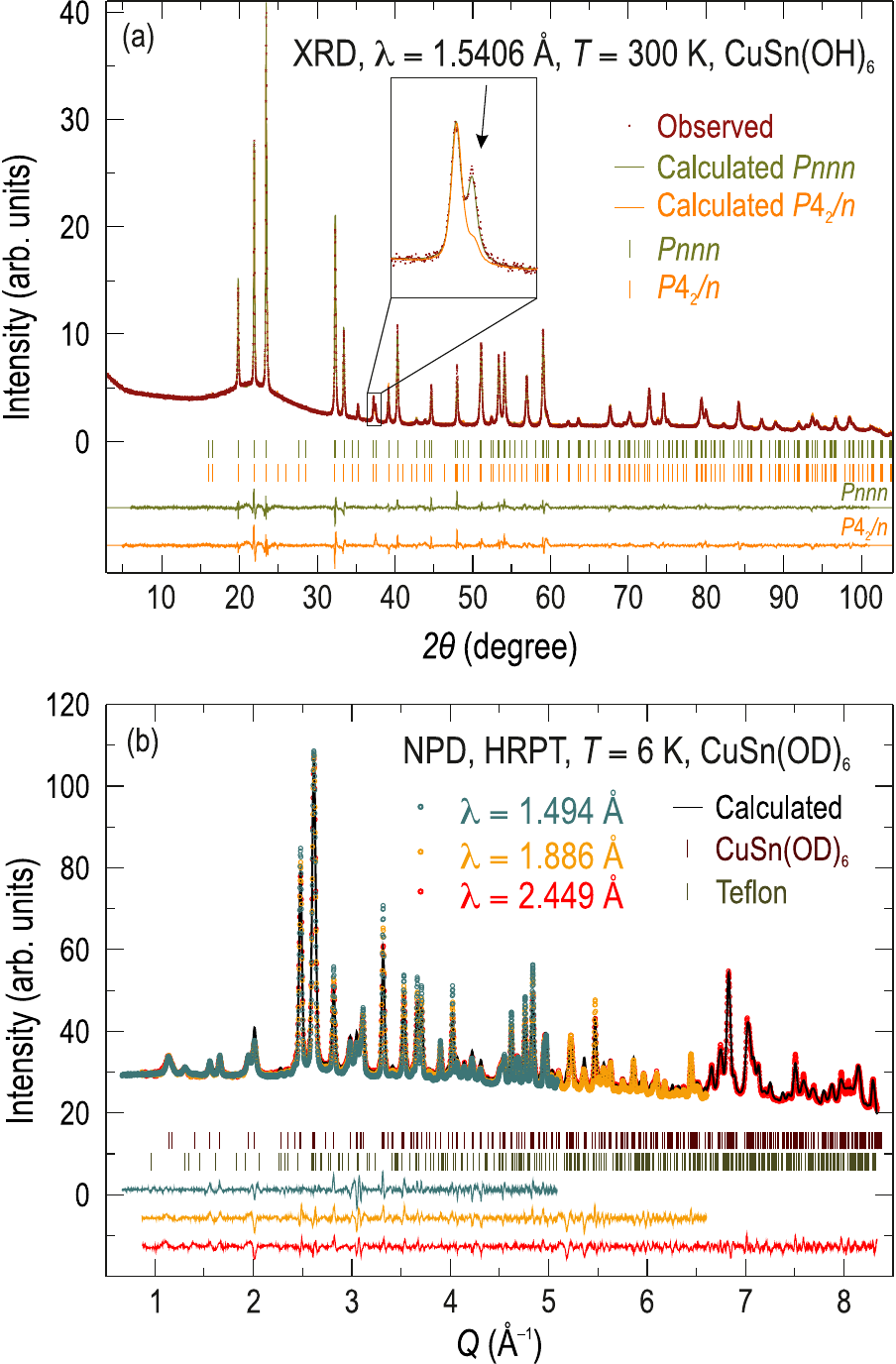}
	\caption{(a)~Scattered (a)~x-ray intensities at 300~K, as a function of 2$\theta$, refined in the tetragonal $P4_{2}/n$ and orthorhombic $Pnnn$ space groups. The inset highlights a region that cannot be adequately described within $P4_{2}/n$. (b) Joint refinement of data at 1.494, 1.886 and 2.449\,\AA\ at HRPT, collected at 6\,K.  Vertical lines denote Bragg positions, and the residuals are shown at the bottom.}
	\label{fig:XRD_NPD}
\end{figure}

Powder x-ray diffraction measurements were performed with two different setups: For Debye-Scherrer-type measurements of the protonated \cusnoh\ sample, a 0.3~mm glass capillary was filled with the sample. The capillary was mounted on a \textsc{Stadi-P} diffractometer (STOE \& Cie), equipped with a Johannsson-type Ge(111) monochromator using Cu-K${\alpha}_{1}$ radiation ($\lambda = 1.5406$~\AA) and a \textsc{Mythen 1K} strip detector (\textsc{Dectris}). Sixteen individual runs were performed at $295(2)$\ K and their intensities were added up for the final evaluation. Collected XRD data on the protonated samples with the refined $Pnnn$ and $P4_{2}/n$ models are shown in Fig.~\ref{fig:XRD_NPD}\,(a).

In Fig.\,\ref{fig:XRD_NPD}\,(b), we also show additional low-temperature NPD data collected at 6~K with three different wavelengths at the HRPT instrument, refined jointly in the $Pnnn$ space group. The 1.494-\AA\ data set is identical to that in Fig.~\ref{fig:neutrons}.

\newpage

\bibliography{main}

\end{document}